\documentclass[fp,twocolumn]{jpsj3}
\usepackage{txfonts}
\usepackage[T1]{fontenc}
\usepackage{ulem}
\usepackage{color}
\usepackage{appendix}
\usepackage{bm}

\title{\textbf{Machine Learning Study on the Flat-Band States Constructed by Molecular-Orbital Representation with Randomness}}

\author{Takumi Kuroda, Tomonari Mizoguchi, Hiromu Araki, and Yasuhiro Hatsugai}
\inst{\textit{Department of Physics, University of Tsukuba, Tsukuba, Ibaraki 305-8571, Japan}} 

\abst{
We study the characteristic probability density distribution of random flat-band models by machine learning. 
The models considered here are constructed on the basis of the molecular-orbital representation, which guarantees the 
existence of macroscopically degenerate zero-energy modes even in the presence of randomness. 
We find that flat-band states are successfully distinguished from conventional extended and localized states, indicating the characteristic feature of the flat-band states.  
We also find that the flat-band states can be detected when the target data 
are defined in a different lattice from the training data, which implies 
the universal feature of the flat-band states constructed by the molecular-orbital representation.
}

\begin{document}
\maketitle
\section{Introduction}
In condensed matter physics, 
lattice models for itinerant electrons exhibiting characteristic dispersions have attracted considerable interest.
A flat-band model is one of their representatives.
Flat-band models have been studied from several points 
of view, mainly for strongly correlated systems, 
such as 
ferromagnetism in the Hubbard model~\cite{Mielke_1991,Tasaki1992}, 
geometrically frustrated antiferromagnets~\cite{Schulenburg2002,Zhitomirsky2004},
topological phases~\cite{Katsura2010,Tang2011,Sun2011,Neupert2011}, Wigner crystals~\cite{Wu2007}, and quantum many-body scars~\cite{Hart2020,Kuno2020,Kuno2021}. 

Recently, another aspect of flat-band systems has gained considerable attention, 
that is, the localized nature of the flat-bands.
Namely, an arbitrary linear combination over a flat-band is also an eigenstate of the Hamiltonian;
thus, one can make a localized eigenstate 
by taking a linear combination from over the entire Brillouin zone.
Such localized eigenstates have long been explored theoretically~\cite{Sutherland1986,Vidal1998,Zhitomirsky2004}
and have indeed been observed experimentally in artificial materials 
such as photonic crystals~\cite{Mukherjee2015,Leykam2018,Leykam2018_2,Liqin2020}.

Moreover, the localization of single-particle eigenstates in random electron systems, i.e., Anderson localization~\cite{Anderson1958}, 
has been extensively studied in condensed matter physics.
Thus, it is natural to ask 
what is characteristic when combining these two ingredients, i.e., flat-bands and randomness.
Several previous works have addressed this issue~\cite{Goda2006,Nishino2007,Chalker2010,Shukla2018,Shukla2018_2,Bilitewski2018} 
by considering models with random potential,
which breaks the exact flat-band.

In this paper, 
we consider yet another class of flat-band models with randomness, 
which we refer to as random molecular-orbital (MO) models~\cite{Kohda2017,Hatsugai2021,Mizoguchi2019}.
The model construction is based on the MO representation, which has been developed to describe generic flat-band models~\cite{Hatsugai2011,Hatsugai2015,Mizoguchi2019,Mizoguchi2020}.
In this class of models, we can introduce randomness while keeping the macroscopically degenerate zero modes. 
In the clean limit, these zero modes are reduced to the flat-band, 
so we simply call the degenerate zero modes the flat-band in what follows.
In this class of models, we study the characteristic wave functions of the flat-band states.
To this end, we perform the supervised machine learning using a convolutional neural network (CNN)
to distinguish the wave functions of the localized, extended, 
and flat-band states.
Machine learning is nowadays 
widely applied to various problems of condensed matter physics. 
Phase classification based on 
single-particle wave functions is one of its typical uses.
Examples include disordered systems~\cite{Ohtsuki2016,Ohtsuki2017,Mano2017}, 
topological phases~\cite{Yoshioka2018,Sun2018,Araki2019,Mano2019}, 
and non-Hermitian systems~\cite{Narayan2021,Araki2021} (see Ref.~41 for a review). 
These studies motivate us to employ CNN for random flat-band models. 

In this work, before presenting the results on machine learning, we address 
the inverse participation ratio (IPR) of flat-band states to analyze their scaling properties.
We find that the IPR shows 
the same system-size dependence as the extended states. 
This indicates that the flat-band states behave as the extended states in the thermodynamic limit. 
Then, we perform phase classification based on supervised learning
using a CNN.
We find that the flat-band states can be distinguished from localized and extended states, 
and that the flat-band states of random MO models have some common features
that are not dependent on microscopic lattice structures.

The rest of this paper is organized as follows.
In Sect~\ref{sec:Model}, we define the MO models used in this study and explain their features in detail. 
In Sect~\ref{sec:IPR}, we show the IPR for the zero modes and discuss its finite-size scaling.
In Sect~\ref{sec:ML}, 
we explain how we apply machine learning and present the results.
Finally, in Sect~\ref{sec:summary}, we present a summary of this paper.
The appendix is devoted to details of the two-dimensional (2D) symplectic model,
from which we extract data of the extended states.

\section{Molecular-orbital representation with randomness \label{sec:Model}}
Our main targets of this study are the
random MO Kagome and MO checkerboard lattice models,
for both of which we consider spinless fermions.

\begin{figure}[b]
 \begin{minipage}{0.495\hsize}
  (a)
    \begin{center}
    \includegraphics[width=0.99\linewidth]{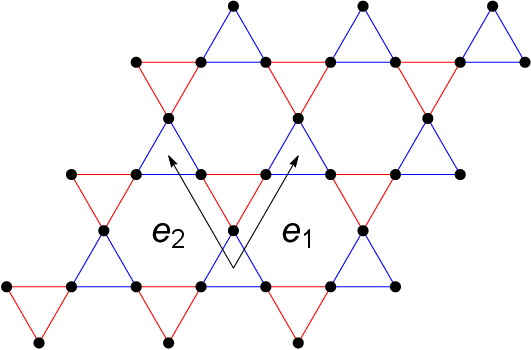}
    \label{fig:km}
    \end{center}
    \end{minipage}
    \begin{minipage}{0.495\hsize}
     (b)
       \begin{center}
     \includegraphics[width=0.75\linewidth]{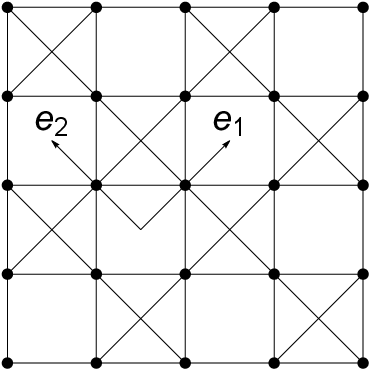}
     \label{fig:ch}
      \end{center}
     \end{minipage}
     \caption{(a) Kagome lattice and (b) checkerboard lattice.
     $\bm{e}_1$ and $\bm{e}_2$ represent the lattice vectors. (Color online)}
     \label{fig:lattice}
\end{figure}

We first elucidate how macroscopically degenerate zero modes appear,
taking the MO Kagome model as an example.
The Kagome lattice has three sublattice degrees of freedom, labeled $j=1,2,3$ [Fig.~\ref{fig:lattice}(a)].
Suppose that the system contains $L \times L$ unit cells
and that the periodic boundary condition is imposed. 
On this lattice, we consider the dimensionless Hamiltonian described as
\begin{align}
    H &= \sum_{\mathbf{r}} C_B^\dagger(\mathbf{r})C_B(\mathbf{r}) + C_R^\dagger(\mathbf{r})C_R(\mathbf{r}), \label{eq:Ham_Kagome}
\end{align}
where
\begin{align}
    C_B(\mathbf{r}) \equiv& \sum_{j=1}^3 \lambda_{B,j}(\mathbf{r})c_j(\mathbf{r}), \notag \\ 
    C_R(\mathbf{r}) \equiv& \sum_{j=1}^3 \lambda_{R,j}(\mathbf{r})c_j(\mathbf{r}+\mathbf{e}_j). \label{eq:MO_Kagome}
\end{align}
Here, $c_j(\mathbf{r})$ is the annihilation operator of a 
fermion at site $j$ in unit cell $\mathbf{r}$,
and $\lambda_{B,j}(\mathbf{r})$ ($\lambda_{R,j}(\mathbf{r})$) 
is a random weight on a blue (red) triangle in Fig.~\ref{fig:lattice}(a).
Note that $C_{B/R}(\mathbf{r})$ is called the MO because
it consists of a linear combination of atomic orbitals (AOs)
within a finite range.
Also note that we have used the notation $\mathbf{e}_3 = 0$ in Eq.~(\ref{eq:MO_Kagome}).

The randomness of the system is introduced via the parameter $\lambda_{B/R,j}(\mathbf{r})$. 
Here, we consider the case where $\lambda_{B,j}(\mathbf{r})$ and $\lambda_{R,j}(\mathbf{r})$ obey the following distribution:
\begin{align}
\lambda_{B/R,j}(\mathbf{r}) = 1 + D^2_{B/R,j}(\mathbf{r}), 
\end{align}
where $D_{j, B/R}(\mathbf{r})$ is a random variable 
obeying the Gaussian distribution whose standard deviation is $v$ and mean is zero.
Note that $\lambda_{B/R,j}$ is positive.

\begin{figure}[t]
 \begin{minipage}{0.495\hsize}
 (a)
  \begin{center}
   \includegraphics[width=\hsize]{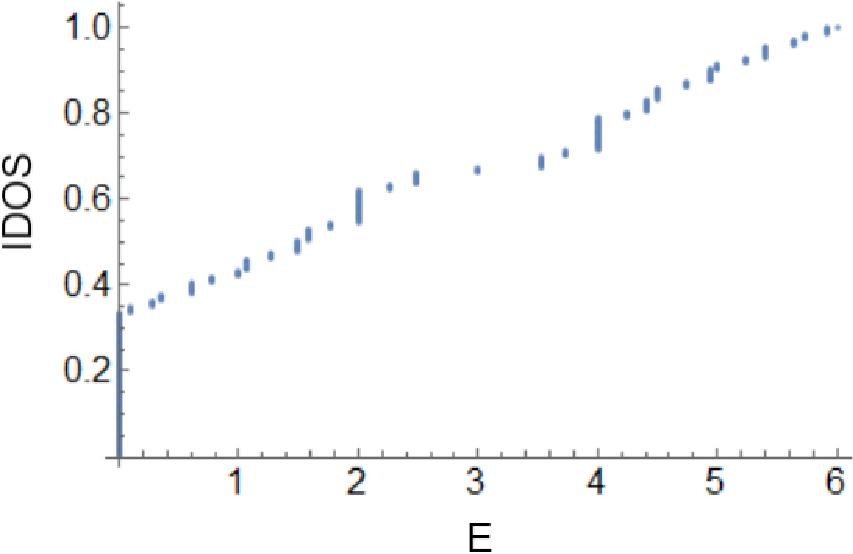}
  \end{center}
 \end{minipage}
 \begin{minipage}{0.495\hsize}
 (b)
  \begin{center}
   \includegraphics[width=\hsize]{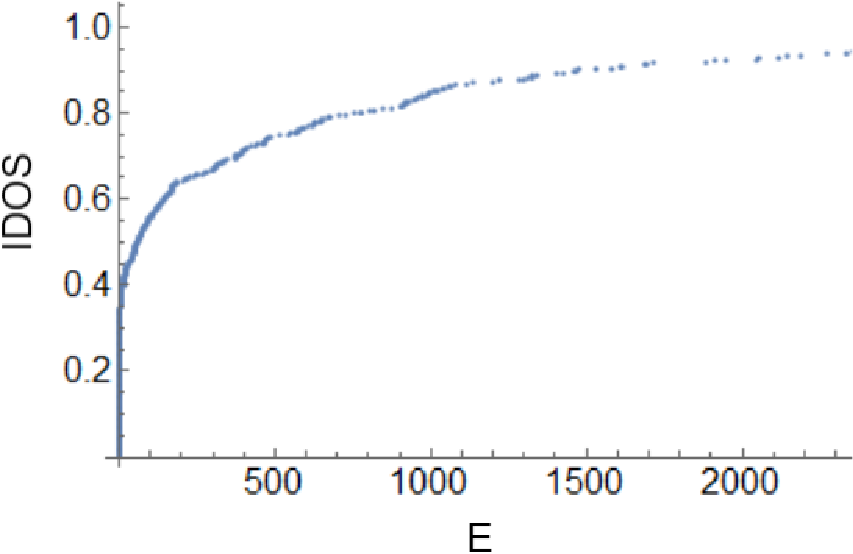}
  \end{center}
 \end{minipage}
 \\
 \begin{minipage}{0.495\hsize}
 (c)
  \begin{center}
   \includegraphics[width=0.9\hsize]{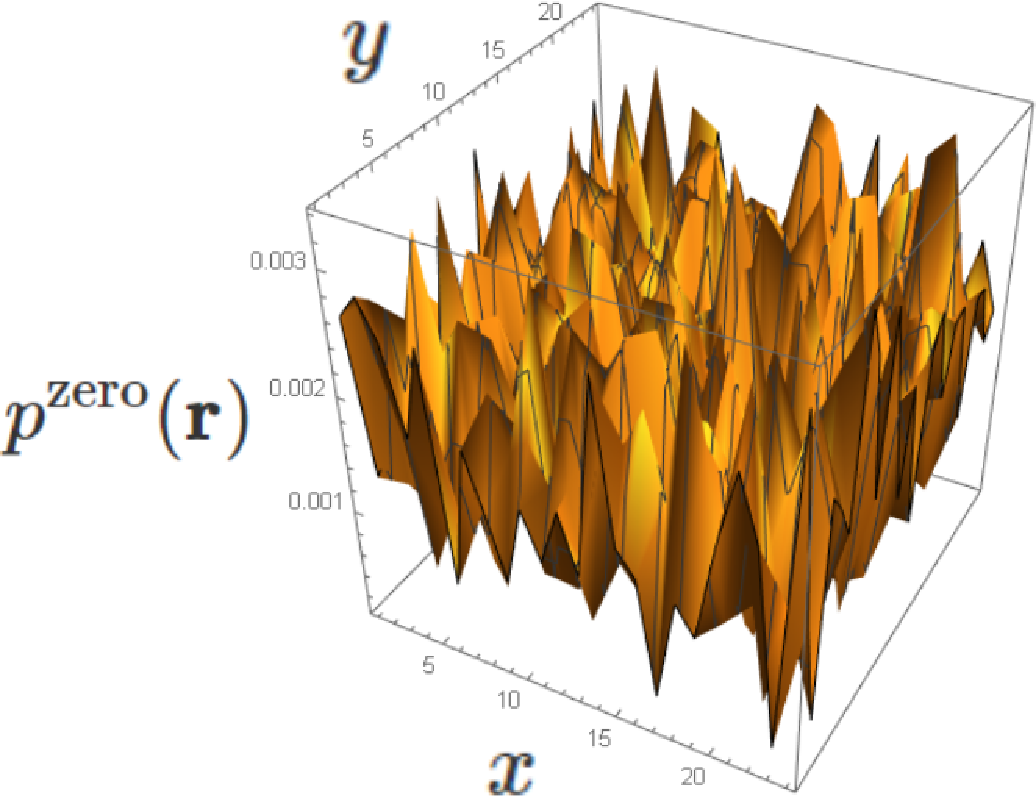}
  \end{center}
 \end{minipage}
 \begin{minipage}{0.495\hsize}
 (d)
  \begin{center}
   \includegraphics[width=0.8\hsize]{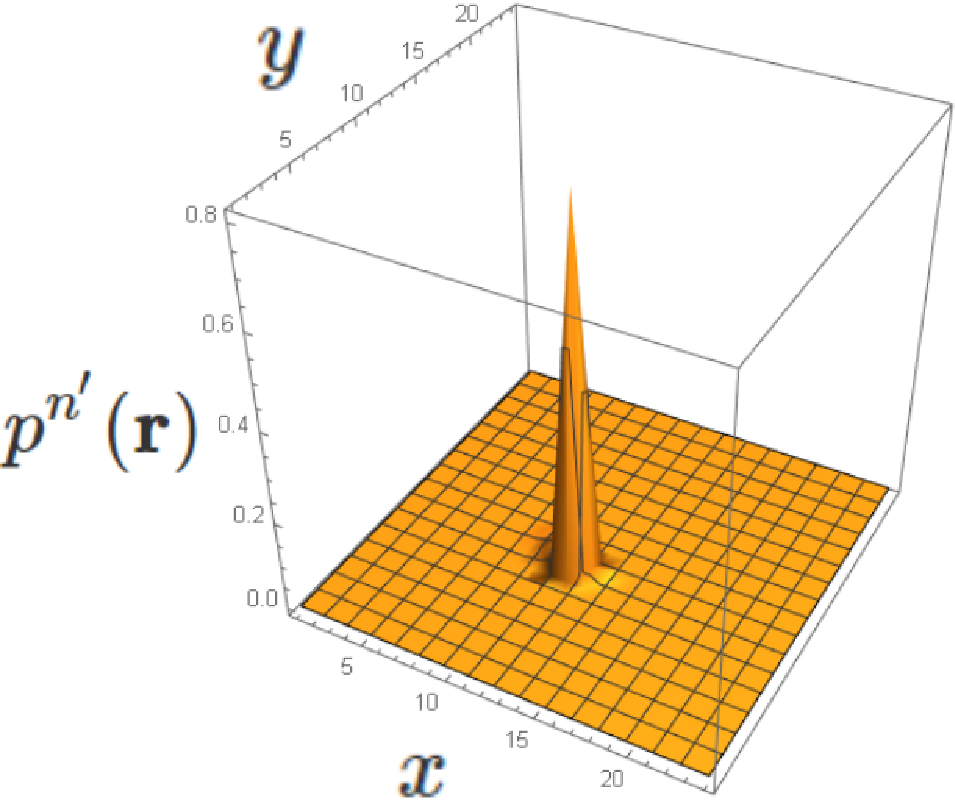}
  \end{center}
 \end{minipage}
 \caption{(a, b) IDOS ($L=24$) in Kagome model. 
 (a) Clean model ($v=0$) and (b) Disordered model ($v=3$).
 (c, d) Probability density in Kagome model ($L=24$), where $x$ and $y$ are defined as $\mathbf{r}=x\bm{e}_1+y\bm{e}_2$. 
 (c) Flat-band states for $v = 1$ and (d) Localized states for $v = 3$. (Color online)}
 \label{fig:kp}
\end{figure}

The origin of the zero modes can be understood from the matrix representation of the Hamiltonian. 
Let $\bm{C}$ be the $2L^2$-component column vector obtained by aligning all the MOs and $\bm{c}$ be the $3 L^2$-component column vector obtained by aligning all the AOs.
They are related to each other through the $2L^2 \times 3 L^2$ matrix
$\Psi^\dagger$ as
\begin{align}
\bm{C} = \Psi^\dagger \bm{c},
\end{align}
where the matrix elements of $\Psi^\dagger$ are determined from Eq.~(\ref{eq:MO_Kagome}).
Then, the Hamiltonian of Eq.~(\ref{eq:Ham_Kagome}) can be written as
\begin{align}
H = \bm{C}^\dagger  \bm{C} =  \bm{c}^\dagger \mathcal{H} \bm{c},
\end{align}
with 
\begin{align}
\mathcal{H} = \Psi \Psi^\dagger \label{eq:Ham_matrix}
\end{align}
being a $3L^2 \times 3L^2$ matrix.
The single-particle eigenstates are obtained as the eigenvectors of the Hamiltonian matrix $\mathcal{H}$. 
As $\Psi^\dagger$ is a $2L^2 \times 3 L^2$ matrix, there exist a set of vectors, $\{ \bm{\varphi}_n | n= 1, \cdots L^2\}$, that satisfies $\Psi^\dagger \bm{\varphi}_n = 0$.
Then, from Eq.~(\ref{eq:Ham_matrix}), one finds that 
the vectors $\bm{\varphi}_n$ $(n=1, \cdots L^2)$ are the zero-energy eigenvectors of $\mathcal{H}$.
This means that the Hamiltonian constructed in this way is guaranteed to host (at least) $L^2$ zero modes, 
even in the presence of randomness in $\lambda_{B/R,j}(\mathbf{r})$.
This can be numerically confirmed by plotting the integrated density of states (IDOS) for the energy as shown in Figs.~\ref{fig:kp}(a) and \ref{fig:kp}(b).

Our interest lies in the wave functions, or probability density distribution, of the eigenstates of this model.
In Figs.~\ref{fig:kp}(c) and \ref{fig:kp}(d), 
we plot the probability density summed over each unit cell for the zero modes
\begin{align}
    p^{\rm zero}(\mathbf{r})=\sum_j p^{\rm zero}_j(\mathbf{r})=\sum_j\left(\frac{1}{L^2}\sum_{n=1}^{L^2}|\varphi^{\rm zero}_{j,n}(\mathbf{r})|^2\right)
    \label{PD}
\end{align}
and the finite-energy mode
\begin{align}
    p^{n^\prime}(\mathbf{r})=\sum_j p_j^{n^\prime}(\mathbf{r})=\sum_j |\varphi_j^{n^\prime}(\mathbf{r})|^2,
    \label{PD2}
\end{align}
respectively.
Here, we write the orthonormal wave functions of the zero modes as $\varphi_{j,n}^{\rm zero}(\mathbf{r})$ with $n=1,\cdots, L^2$, and those of the finite-energy mode as $\varphi_j^{n^\prime}(\mathbf{r})$ with $n^\prime=1,\cdots, 2L^2$.
As the zero modes are $L^2$-fold degenerate, the probability density in Eq.~(\ref{PD}) is traced over all the degenerate states to become invariant under unitary transformation.
Note that one can calculate the probability density by using the inverse matrix of the overlap matrix $\Psi^\dagger \Psi$, instead of by diagonalizing $\mathcal{H}$ directly~\cite{Hatsugai2021}.
We see that the probability density of the flat-band state in Fig.~\ref{fig:kp}(c) has a spiky spatial distribution, which is clearly distinct from the localized state in Fig.~\ref{fig:kp}(d) (see below).
Although extended states of random systems also have a spiky spatial distribution [see, e.g., Fig. A.1(a)], they are distinguished from the characteristic distribution of flat-band states by machine learning, as we will show in Sect~\ref{sec:ML}.
For the finite-energy state [Fig.~\ref{fig:kp}(d)], the localization length of the finite-energy state (for a sufficiently large value of $v$) is clearly smaller than the system size, which coincides with the scaling theory~\cite{Abrahams1979}, indicating that generic eigenstates of 2D systems 
are localized owing to randomness.

Thus far, we have explained the Kagome model. Another lattice model, namely, the checkerboard model, can be constructed 
in the same way as the Kagome model.
In this lattice, there are two sublattice degrees of freedom, $j=1,2$.
In this model, we consider the Hamiltonian
\begin{align}
    H = \sum_{\mathbf{r}} C^\dagger(\mathbf{r})C(\mathbf{r}),
\end{align}
where
\begin{align}
    C(\mathbf{r}) \equiv 
    \lambda_{1}(\mathbf{r})c_1(\mathbf{r})+\lambda_{2}(\mathbf{r})c_2(\mathbf{r})+ \lambda_{3}(\mathbf{r})c_1(\mathbf{r}+\mathbf{e}_1)+\lambda_{4}(\mathbf{r})c_2(\mathbf{r}+\mathbf{e}_2), \notag \\
\end{align}
and 
\begin{align}
\lambda_{\ell}(\mathbf{r}) = 1 + D_{\ell}^2 \hspace{1mm}(\ell = 1,2,3,4).
\end{align}
Again, $D_{\ell}$ obeys the Gaussian distribution. 
Then, the model possesses (at least) $2\times L^2-L^2=L^2$ zero modes, as is the case of the Kagome model, since it consists of $2\times L^2$ AOs and $L^2$ MOs.

\section{IPR for zero modes \label{sec:IPR}}

\begin{figure}[t]
  \begin{center}
   \includegraphics[width=0.75\hsize]{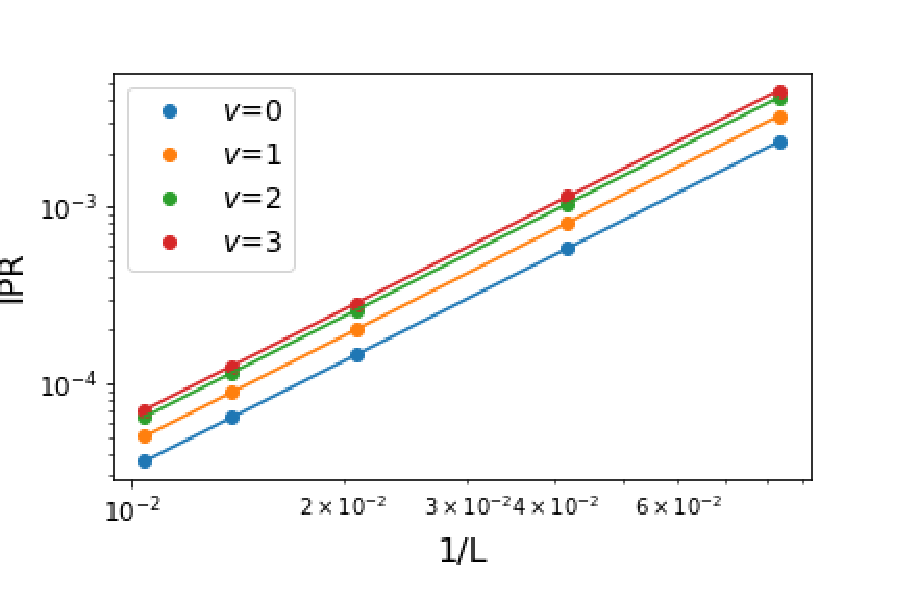}
  \end{center}
 \caption{IPR versus $1/L$ for each standard deviation, where averages over five samples are taken. The dots are the plots at $1/L=1/12,1/24,1/48,1/72$, and $1/96$ and the lines are the fitting functions ,with the parameters shown in Table~\ref{t1}. (Color online)}
 \label{fig:ipr}
\end{figure}

\begin{table}[t]
\begin{center}
\begin{tabular}{c|cc}
\hline
\multicolumn{1}{c}{} & \multicolumn{1}{c}{$a$} & \multicolumn{1}{c}{$b$} \\
\hline
\verb  $v=0$ & $2.000$ & $0.3334$ \\
\verb  $v=1$ & $2.006$ & $0.4736$ \\
\verb  $v=2$ & $2.000$ & $0.5959$ \\
\verb  $v=3$ & $2.002$ & $0.6566$ \\
\hline
\end{tabular}
\end{center}
\caption{Fitting parameters of $\text{IPR}=b/L^a$ for each standard deviation.}
\label{t1}
\end{table}

Before proceeding to the machine learning study, 
we ,derive the scaling properties of the zero modes of random MO models. 
Specifically, we refer to the finite-size scaling of their IPR.
To ensure that the IPR is well defined even for macroscopically degenerate states, 
we take a trace over all the degenerate zero modes.
By doing so, it is expected that the IPR will behave as an extended state 
because the degenerate zero modes appear as the complementary space of the finite-energy modes~\cite{Hatsugai2021}.
To be more specific, each finite-energy mode is localized, 
but the set of macroscopic localized states has the same probability density distribution as the extended states.
Thus, as the eigenstates of $H$ form a complete set, 
the probability density distribution of the set of zero modes is given by subtracting that for the finite-energy modes from the uniform distribution.
Hence, the probability density distribution for the set of zero modes 
also behaves as an extended state.

We define the IPR 
for the entire set of flat-band states as 
\begin{align}
 \text{IPR} = \sum_{\mathbf{r}} \sum_{j=1}^3 [p_j^{\rm zero}(\mathbf{r})]^2
\end{align}
in the Kagome model. 
We calculate the IPR while varying the system size
from $L=12$ to $L=96$ 
for each standard deviation $v=0,1,2,3$.
The result is shown in Fig.~\ref{fig:ipr}.
Here, arithmetic means of five samples are taken.
We see that, for all values of $v$, the IPR is a power function of $1/L$.
To find the exponent, we fit the data by the function, 
$\text{IPR} = b/L^a$.
The results are summarized in Table~\ref{t1},
where we see that $\text{IPR} \propto 1/L^2$.
This indicates that the flat-band states
obeys the same scaling law as the extended states.

\section{Machine learning of flat-band states \label{sec:ML}}
In this section, we describe the results of the machine learning study of the wave functions of random MO models.

\subsection{Method}
\begin{figure}[t]
  \begin{center}
   \includegraphics[width=\hsize]{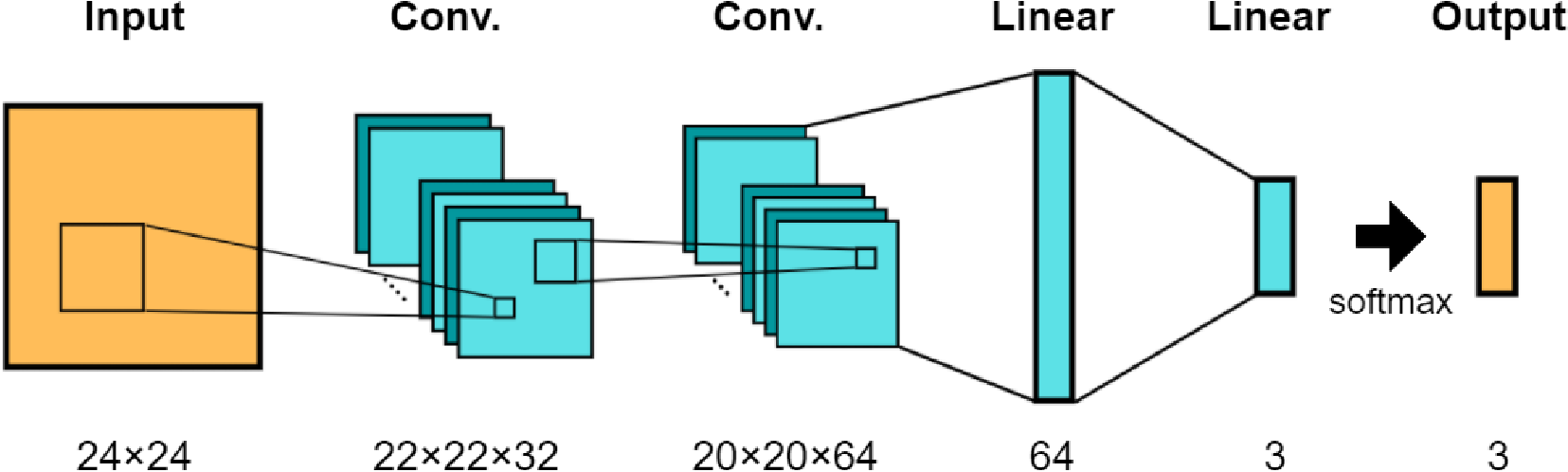}
  \end{center}
 \caption{Layer structure of the network. ``Conv.'' and ``Linear'' mean convolutional and affine layers respectively, and the numbers represent the dimensions of each data. (Color online)}
 \label{fig:mls}
\end{figure}

We employ machine learning using a CNN
to characterize the localized, extended, and flat-band states.
Here, a neural-network-based algorithm is implemented by using the open-source library PyTorch~\cite{PyTorch}. 
A schematic figure of the architecture of the CNN is shown in Fig.~\ref{fig:mls}.
It consists of two convolutional layers and two linear layers. 
The numbers of filters comprising the convolutional layers are 32 and 64. 
The kernel size, stride, and padding of both convolutional layers are $3\times3$, 1, and 0, respectively.
The first linear layer transforms the output of the previous layer into a 64-component vector and the second one transforms it into a 3-component vector.
All the activation functions are ReLU functions.
For input $x$, the ReLU function returns $x$ if $x>0$, otherwise it returns $0$.
After the second convolutional layer, a max pooling layer whose kernel size is $2\times2$ is there.
To prevent overfitting, a dropout layers are inserted after the pooling layer and the first linear layer. 
The probability of dropout is set to 0.25 for the first one and 0.5 for the other.

\begin{table}[t]
\begin{center}
\begin{tabular}{c|ccc}
\hline
\multicolumn{1}{c}{} & \multicolumn{1}{c}{Flat-band} & \multicolumn{1}{c}{Localized} & \multicolumn{1}{c}{Extended} \\
\hline
(A) & K(500), CB(500) & K(250), CB(250), 2D(500) & 2D(1000) \\
(B) & K(1000) & K(500), 2D(500) & 2D(1000) \\
\hline
\end{tabular}
\end{center}
\caption{Breakdown of the training data for machines (A) and (B). ``K'', ``CB'', and ``2D'' are abbreviations of ``Kagome model'', ``checkerboard model'', and ``2D symplectic model'', respectively, and the numbers in parentheses indicate the number of data.}
\label{t2}
\end{table}

We prepare the training data of the above three states for the supervised learning as follows.
The flat-band and localized states are extracted from 
the MO Kagome and MO checkerboard models.
For a given Hamiltonian from a random MO model, 
we obtain the probability density for each unit cell, 
averaged over the $L^2$ zero modes, which is invariant under unitary transformation, as in Eq.~(\ref{PD}).
In addition to the flat-band states, 
we also take samples of the localized states by picking up one of the finite-energy modes randomly as in Eq.~(\ref{PD2}).
Note that the disorder strength is chosen to be sufficiently large, as shown later.
Each $p^{\rm zero}(\mathbf{r})$ and $p^{n^\prime}(\mathbf{r})$ 
is $L \times L$ data, adopted as training data.
Here, we set $L=24$ and normalize the input data so that the average of all the components is $0$ and the standard deviation is $1$.
We construct two types of machines:
\begin{enumerate}
\item[(A)] For flat-band states, we take 500 samples from the MO Kagome model and 500 samples from the MO checkerboard model (thus, 1000 data in total), with $v$ dividing into equal parts in $v \in [1.0,3.0]$.
For localized states, we take 250 samples each from the MO Kagome and checkerboard models. 
\item[(B)] For flat-band states, we take 1000 samples from the MO Kagome model in the same $v$ as for machine (A).
For localized states, we take 500 samples from the MO Kagome model.
\end{enumerate}

In addition to data of the flat-band and localized states, 
we also train each machine to distinguish the extended state. 
However, the extended states cannot be obtained from the MO Kagome/checkerboard model
in the presence of finite randomness.
To overcome this difficulty, we employ another model, namely, the 2D symplectic model, which is known to host the extended states. 
Details of the model are shown in Appendix. 
In this model, there is a critical value of disorder, $W_c \sim 6.2$ (see Appendix for the definition of $W$), below which the extended states survive at $E\sim 0$ (i.e., the band center).
Hence, we calculate the probability density (per unit cell)
for the mode with $E \sim 0$
while dividing the strength of randomness $W$ into equal parts in $W\in [0.0,3.1]$,
and adopt the data thus obtained as the training data for the extended state.
We take 1000 independent samples. 
In addition, we take 500 samples for the localized states 
with $W\in [9.3,12.4]$ (i.e., $W > W_c$).
Here, the size of each data is $24 \times 24$ and all the data are normalized in the same manner.

From these procedures, we prepare 1000 samples for each of the localized, extended, and flat-band states. 
The preparation of training data described thus far is summarized in Table~\ref{t2}.
Note that
the training data are collected from multiple models, rather than a single model.
Therefore, the present CNN is expected to provide a suitable platform to test
whether the lattice-structure-independent features can be extracted by machine learning.

Using the trained CNN, we perform phase classifications 
among the localized, extended, and flat-band states. 
The target data are taken from the MO Kagome and MO checkerboard models. 
We prepare the data of the probability density in the same manner as for the training data,
but we do not restrict $v$ to large value.
When inputting one data, we obtain an output in the form of a three-component vector, $\bm{P} = (P_{F},P_{L},P_{E})$, with $0\leq P_{F},P_{L},P_{E} \leq 1$.
The components satisfy 
\begin{align}
P_{F} + P_{L} + P_{E} =1.
\end{align}
As such, each component can be interpreted as the probability that the input data belongs to each phase. 

\begin{figure}[t]
 \begin{minipage}{0.495\hsize}
 (a)
  \begin{center}
   \includegraphics[width=\hsize]{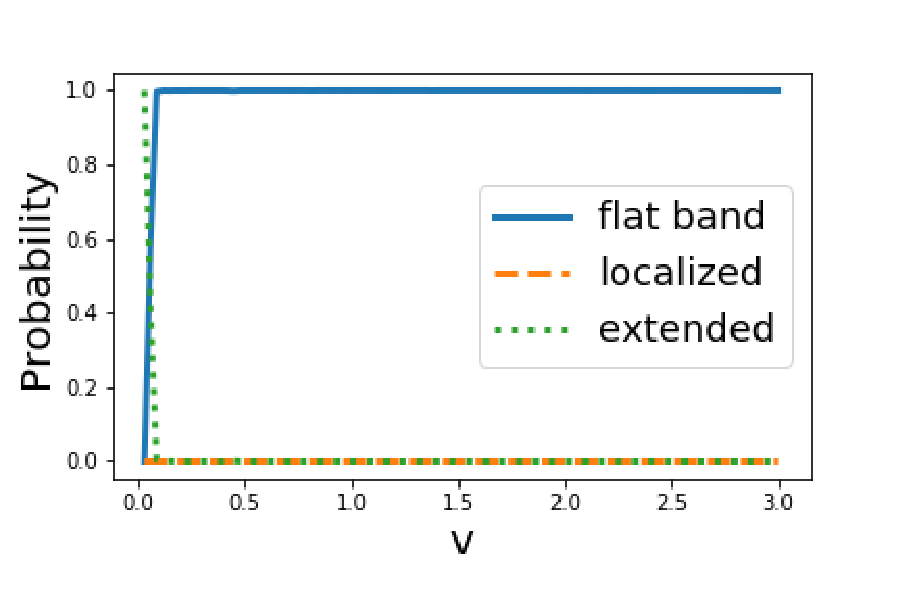}
  \end{center}
  \label{fig:fb1}
 \end{minipage}
 \begin{minipage}{0.495\hsize}
 (b)
  \begin{center}
   \includegraphics[width=\hsize]{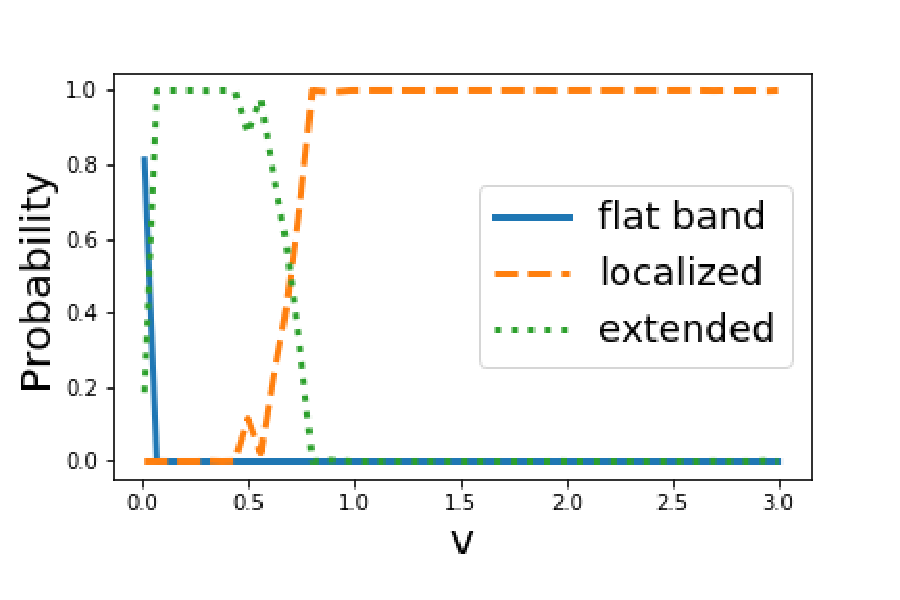}
  \end{center}
  \label{fig:fb2}
 \end{minipage}
 \\
 \begin{minipage}{0.495\hsize}
 (c)
  \begin{center}
   \includegraphics[width=\hsize]{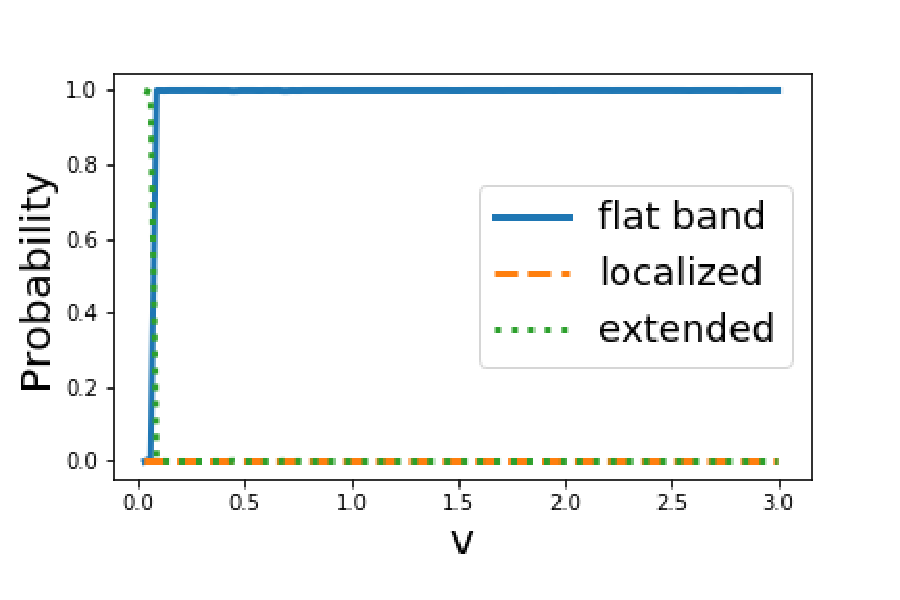}
  \end{center}
  \label{fig:kc}
 \end{minipage}
 \begin{minipage}{0.495\hsize}
 (d)
  \begin{center}
   \includegraphics[width=\hsize]{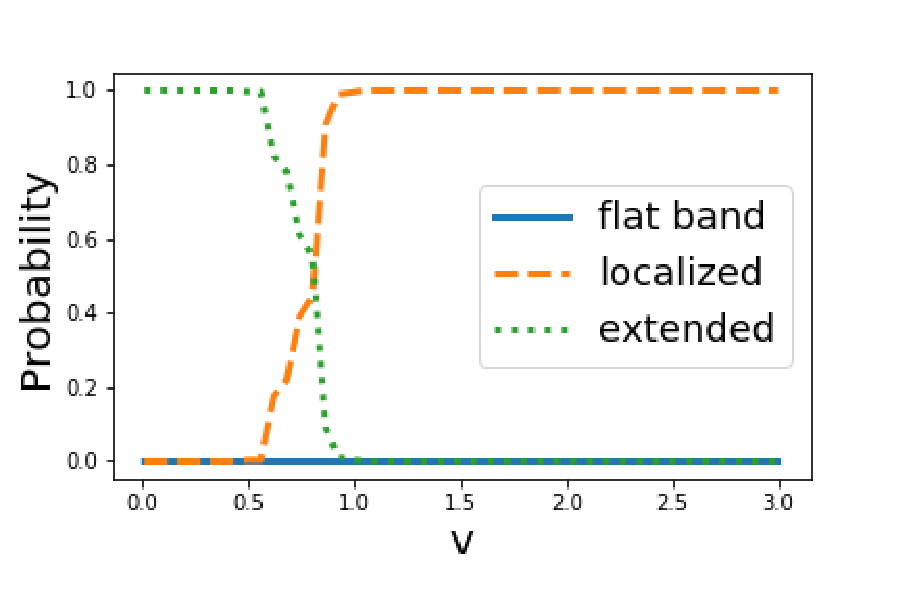}
  \end{center}
  \label{fig:kl}
 \end{minipage}
 \caption{Evaluations of $\mathbf{P}$ by machines 
where averages over five samples are taken.
 (a) Flat-band states in the Kagome model for machine (A). 
 (b) Finite-energy modes in the Kagome model for machine (A).
 (c) Flat-band states in the checkerboard model for machine (B). 
 (d) Finite-energy modes in the checkerboard model for machine (B). (Color online)}
 \label{fig:SCL}
\end{figure}

\begin{figure}[t]
 \begin{minipage}{0.32\hsize}
  \begin{center}
   \includegraphics[width=\hsize]{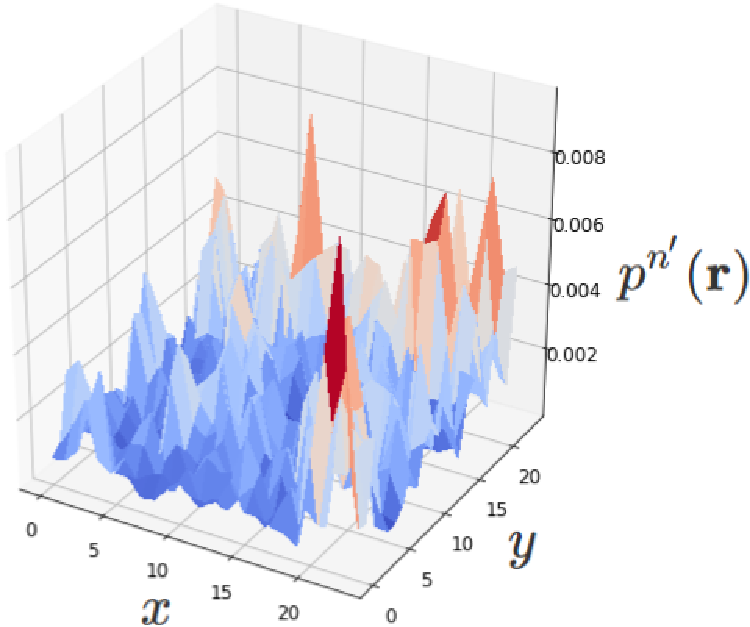}
  \end{center}
 \end{minipage}
 \begin{minipage}{0.32\hsize}
  \begin{center}
   \includegraphics[width=\hsize]{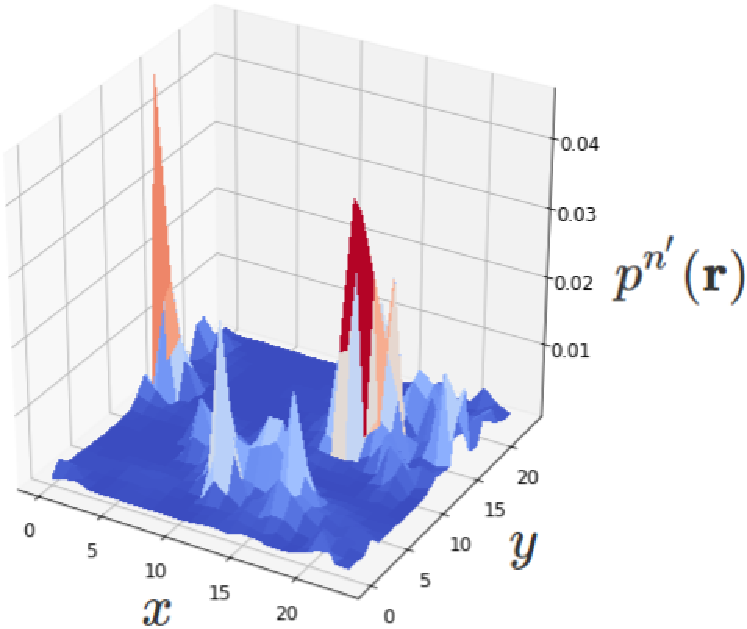}
  \end{center}
 \end{minipage}
 \begin{minipage}{0.32\hsize}
  \begin{center}
   \includegraphics[width=\hsize]{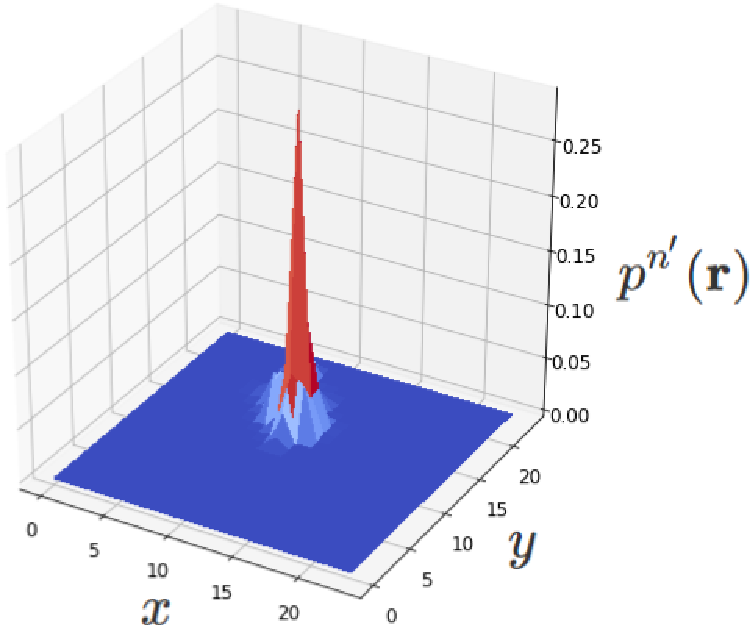}
  \end{center}
 \end{minipage}
 \caption{Probability densities of the finite-modes in the Kagome model, where $x$ and $y$ are defined as $\mathbf{r}=x\bm{e}_1+y\bm{e}_2$.
 From left to right, the standard deviations are $v=0.3$, $v=0.6$, and $v=1.2$. (Color online)}
 \label{fig:pd_nz}
\end{figure}

\subsection{Results \label{sec:results}}
We first present the results where the input data are taken from 
the MO Kagome model and the phase classification is performed 
using machine (A)
to examine 
whether the CNN can distinguish the flat-band states from the other states. 
In Figs.~\ref{fig:SCL}(a) and \ref{fig:SCL}(b),
we show the probability obtained by machine learning
for the zero modes (i.e., the flat-band states) and finite-energy modes, respectively.
Here, we adopt the states with the $L^2/2$th largest energy of all the finite-energy modes for each Hamiltonian as the evaluation data.
The horizontal axis denotes the standard deviation of the randomness $v$. 
In Fig.~\ref{fig:SCL}(a), we find that machine (A) successfully distinguishes
flat-band states from the other states for the entire $v$ range.
This indicates that the flat-band state has a characteristic 
probability density, which is distinct from those of the localized and extended states. 
For the finite-energy modes shown in Fig.~\ref{fig:SCL}(b), we see that 
the phase classification between the localized and extended states fails for small $v$ ($\lesssim 1$).
This is because the localization length is larger than the system size;
the typical probability density distributions for the finite-energy modes are shown in 
Fig.~\ref{fig:pd_nz}.

We next present the results where the input data are taken from 
the MO checkerboard model and the phase classification is performed 
using machine (B), to examine the similarity/difference between
the flat-band states of different lattice structures.
In Figs.~\ref{fig:SCL}(c) and \ref{fig:SCL}(d),
we show the probability obtained by machine learning
for the zero modes (i.e., the flat-band states) and finite-energy modes, respectively.
Here, we adopt the states with the $L^2/4$th largest energy of all the finite-energy modes for each Hamiltonian as the evaluation data.
The result for the finite-energy modes [Fig.~\ref{fig:SCL}(d)]
is qualitatively the same as that in Fig.~\ref{fig:SCL}(b).
For the flat-band states [Fig.~\ref{fig:SCL}(c)], we find that 
the CNN successfully distinguishes flat-band states from the other states except for small $v$ ($v \lesssim 0.06$),
despite the fact that the training data are 
taken from different lattice models from the input data (i.e., the MO Kagome model).
This result indicates that the flat-band states 
of the random MO models have universal features
that are not dependent on microscopic lattice structures.

\section{Summary \label{sec:summary}}
We have used machine learning to study the characteristic 
features of flat-band wave functions of the tight-binding models constructed by the MO representation with randomness.
As demonstrated by IPR analysis, the flat-band states in the random MO model show the scaling behavior of the extended state, 
as they are the ``complement'' of the macroscopic localized (finite-energy) states.
Nevertheless, the flat-band wave functions have unique features 
in the ``submicroscopic'' length scale.
This is demonstrated by a machine learning study,
in which the CNN distinguished the flat-band states from the extended and localized states.
We also reveal that the CNN can distinguish the flat-band states 
even when the training data are taken from different lattices from the input data, indicating the existence of some common features among the flat-band states of random MO models. 

\begin{acknowledgement}
This work is partly supported by JSPS KAKENHI
(Grant Nos.~JP17H06138 and JP20K14371) (T. M.) and JST CREST (Grant No. JPMJCR19T1), Japan.
\end{acknowledgement}

\appendix
\section{2D symplectic model}

\begin{figure}[t]
 \begin{minipage}{0.495\hsize}
 (a)
  \begin{center}
   \includegraphics[width=\hsize]{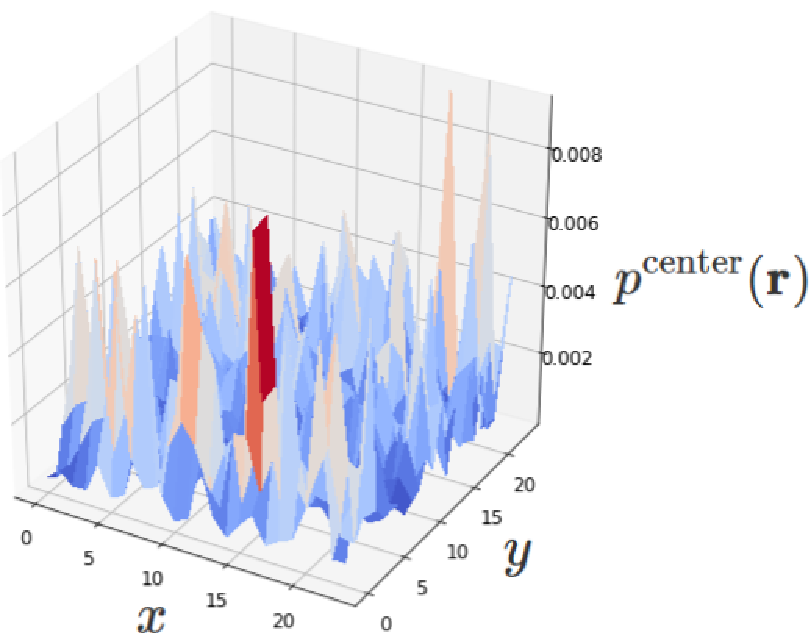}
  \end{center}
 \end{minipage}
 \begin{minipage}{0.495\hsize}
 (b)
  \begin{center}
   \includegraphics[width=\hsize]{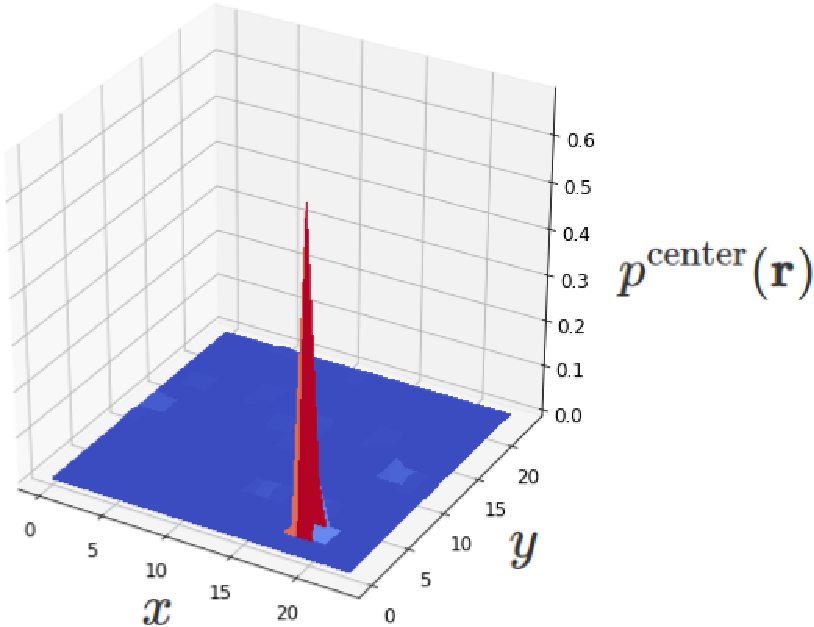}
  \end{center}
 \end{minipage}
 \caption{Probability densities for the states of $E \sim 0$ in the 2D symplectic model, where $x$ and $y$ are defined as $\mathbf{r}=x\bm{e}_1+y\bm{e}_2$. (a) $W<W_c$, and (b) $W>W_c$. (Color online)}
 \label{fig:pd_2d}
\end{figure}

In this appendix, we explain the 2D symplectic model, which we employ
to obtain the data of the extended and localized states.
This model is for $S=1/2$ fermions on a square lattice.
We impose the periodic boundary condition with $L\times L$ unit cells.
The Hamiltonian is described as
\begin{align}
    H=\sum_{i,\sigma}\epsilon_i c_{i,\sigma}^\dagger c_{i,\sigma} - \sum_{\langle i,j\rangle,\sigma,\sigma^\prime}R(i,j)_{\sigma,\sigma^\prime} c_{i,\sigma}^\dagger c_{j,\sigma^\prime}
\end{align}
with
\begin{align}
    R(i,j)=
    \begin{pmatrix}
    e^{i\alpha_{i,j}}\cos{\beta_{i,j}} & e^{i\gamma_{i,j}}\sin{\beta_{i,j}}\\ 
    -e^{-i\gamma_{i,j}}\sin{\beta_{i,j}} & e^{-i\alpha_{i,j}}\cos{\beta_{i,j}}
    \end{pmatrix}.
\end{align}
Here, $c_{i,\sigma}$ is the annihilation operator of a fermion whose spin is $\sigma$ ($\sigma=+,-$) at site $i$, and $\epsilon_{i}$ is a random potential energy on site $i$ that is distributed uniformly at the interval $[-W/2,W/2]$.
Parameters $\alpha$ and $\gamma$ obey the uniform distribution
in the range $[0,2\pi)$, and $\beta$ obeys the following distribution:
\begin{align}
    P(\beta)=\begin{cases}
    \sin(2\beta) & (0\leq\beta\leq\pi/2) \\
    0 & \text{otherwise}
  \end{cases}.
\end{align}

Previous studies~\cite{Asada2002,Asada2004}
revealed that,
for the state with $E\sim 0$,
there exists a critical value of $W_c~(\sim 6.2)$ below which the 
state is extended.
Therefore, in this study, we adopt the probability densities
\begin{align}
    p^{\rm center}(\mathbf{r})=\sum_\sigma p_\sigma^{\rm center}(\mathbf{r})=\sum_\sigma |\varphi_\sigma^{\rm center}(\mathbf{r})|^2, \label{eq:PD_2DS}
\end{align}
with the wave function $\varphi_\sigma^{\rm center}(\mathbf{r})$ whose energy is nearest to the band center $E=0$
for $W<W_c$ as training data for the extended state.
We also use the probability density for 
$W > W_c$
as a training data for the localized state.
In Figs.~\ref{fig:pd_2d}(a) and \ref{fig:pd_2d}(b),
we show the probability densities of Eq.~(\ref{eq:PD_2DS})
for $W<W_c$  and $W>W_c$, respectively.

\bibliographystyle{jpsj}
\bibliography{FB.bib}

\begin{thebibliography}{10}

\bibitem{Mielke_1991}
A.~Mielke: J. Phys. A: Mathematical and General {\bfseries 24} (1991) 3311.

\bibitem{Tasaki1992}
H.~Tasaki: Phys. Rev. Lett. {\bfseries 69} (1992) 1608.

\bibitem{Schulenburg2002}
J.~Schulenburg, A.~Honecker, J.~Schnack, J.~Richter, and H.-J. Schmidt: Phys.
  Rev. Lett. {\bfseries 88} (2002) 167207.

\bibitem{Zhitomirsky2004}
M.~E. Zhitomirsky and H.~Tsunetsugu: Phys. Rev. B {\bfseries 70} (2004) 100403.

\bibitem{Katsura2010}
H.~Katsura, I.~Maruyama, A.~Tanaka, and H.~Tasaki: {EPL} {\bfseries 91} (2010)
  57007.

\bibitem{Tang2011}
E.~Tang, J.-W. Mei, and X.-G. Wen: Phys. Rev. Lett. {\bfseries 106} (2011)
  236802.

\bibitem{Sun2011}
K.~Sun, Z.~Gu, H.~Katsura, and S.~Das~Sarma: Phys. Rev. Lett. {\bfseries 106}
  (2011) 236803.

\bibitem{Neupert2011}
T.~Neupert, L.~Santos, C.~Chamon, and C.~Mudry: Phys. Rev. Lett. {\bfseries
  106} (2011) 236804.

\bibitem{Wu2007}
C.~Wu, D.~Bergman, L.~Balents, and S.~Das~Sarma: Phys. Rev. Lett. {\bfseries
  99} (2007) 070401.

\bibitem{Hart2020}
O.~Hart, G.~De~Tomasi, and C.~Castelnovo: Phys. Rev. Res. {\bfseries 2} (2020)
  043267.

\bibitem{Kuno2020}
Y.~Kuno, T.~Mizoguchi, and Y.~Hatsugai: Phys. Rev. B {\bfseries 102} (2020)
  241115.

\bibitem{Kuno2021}
Y.~Kuno, T.~Mizoguchi, and Y.~Hatsugai: Phys. Rev. B {\bfseries 104} (2021)
  085130.

\bibitem{Sutherland1986}
B.~Sutherland: Phys. Rev. B {\bfseries 34} (1986) 5208.

\bibitem{Vidal1998}
J.~Vidal, R.~Mosseri, and B.~Dou\ifmmode~\mbox{\c{c}}\else \c{c}\fi{}ot: Phys.
  Rev. Lett. {\bfseries 81} (1998) 5888.

\bibitem{Mukherjee2015}
S.~Mukherjee, A.~Spracklen, D.~Choudhury, N.~Goldman, P.~\"Ohberg,
  E.~Andersson, and R.~R. Thomson: Phys. Rev. Lett. {\bfseries 114} (2015)
  245504.

\bibitem{Leykam2018}
D.~Leykam and S.~Flach: APL Photonics {\bfseries 3} (2018) 070901.

\bibitem{Leykam2018_2}
D.~Leykam, A.~Andreanov, and S.~Flach: Adv. in Phys. X {\bfseries 3} (2018)
  1473052.

\bibitem{Liqin2020}
L.~Tang, D.~Song, S.~Xia, S.~Xia, J.~Ma, W.~Yan, Y.~Hu, J.~Xu, D.~Leykam, and
  Z.~Chen: Nanophotonics {\bfseries 9} (2020) 1161.

\bibitem{Anderson1958}
P.~W. Anderson: Phys. Rev. {\bfseries 109} (1958) 1492.

\bibitem{Goda2006}
M.~Goda, S.~Nishino, and H.~Matsuda: Phys. Rev. Lett. {\bfseries 96} (2006)
  126401.

\bibitem{Nishino2007}
S.~Nishino, H.~Matsuda, and M.~Goda: J. Phys. Soc. Jpn. {\bfseries 76} (2007)
  024709.

\bibitem{Chalker2010}
J.~T. Chalker, T.~S. Pickles, and P.~Shukla: Phys. Rev. B {\bfseries 82} (2010)
  104209.

\bibitem{Shukla2018}
P.~Shukla: Phys. Rev. B {\bfseries 98} (2018) 054206.

\bibitem{Shukla2018_2}
P.~Shukla: Phys. Rev. B {\bfseries 98} (2018) 184202.

\bibitem{Bilitewski2018}
T.~Bilitewski and R.~Moessner: Phys. Rev. B {\bfseries 98} (2018) 235109.

\bibitem{Kohda2017}
K. Kohda, Master's thesis, University of Tsukuba (in Japanese) (2017).

\bibitem{Hatsugai2021}
Y.~Hatsugai: Ann. Phys. {\bfseries 435} (2021) 168453.

\bibitem{Mizoguchi2019}
T.~Mizoguchi and Y.~Hatsugai: {EPL} {\bfseries 127} (2019) 47001.

\bibitem{Hatsugai2011}
Y.~Hatsugai and I.~Maruyama: {EPL} {\bfseries 95} (2011) 20003.

\bibitem{Hatsugai2015}
Y.~Hatsugai, K.~Shiraishi, and H.~Aoki: New J. Phys. {\bfseries 17} (2015)
  025009.

\bibitem{Mizoguchi2020}
T.~Mizoguchi and Y.~Hatsugai: Phys. Rev. B {\bfseries 101} (2020) 235125.

\bibitem{Ohtsuki2016}
T.~Ohtsuki and T.~Ohtsuki: J. Phys. Soc. Jpn. {\bfseries 85} (2016) 123706.

\bibitem{Ohtsuki2017}
T.~Ohtsuki and T.~Ohtsuki: J. Phys. Soc. Jpn. {\bfseries 86} (2017) 044708.

\bibitem{Mano2017}
T.~Mano and T.~Ohtsuki: J. Phys. Soc. Jpn. {\bfseries 86} (2017) 113704.

\bibitem{Yoshioka2018}
N.~Yoshioka, Y.~Akagi, and H.~Katsura: Phys. Rev. B {\bfseries 97} (2018)
  205110.

\bibitem{Sun2018}
N.~Sun, J.~Yi, P.~Zhang, H.~Shen, and H.~Zhai: Phys. Rev. B {\bfseries 98}
  (2018) 085402.

\bibitem{Araki2019}
H.~Araki, T.~Mizoguchi, and Y.~Hatsugai: Phys. Rev. B {\bfseries 99} (2019)
  085406.

\bibitem{Mano2019}
T.~Mano and T.~Ohtsuki: J. Phys. Soc. Jpn. {\bfseries 88} (2019) 123704.

\bibitem{Narayan2021}
B.~Narayan and A.~Narayan: Phys. Rev. B {\bfseries 103} (2021) 035413.

\bibitem{Araki2021}
H.~Araki, T.~Yoshida, and Y.~Hatsugai: J. Phys. Soc. Jpn. {\bfseries 90} (2021)
  053703.

\bibitem{Abrahams1979}
E.~Abrahams, P.~W. Anderson, D.~C. Licciardello, and T.~V. Ramakrishnan: Phys.
  Rev. Lett. {\bfseries 42} (1979) 673.

\bibitem{PyTorch}
PyTorch: https://pytorch.org.

\bibitem{Asada2002}
Y.~Asada, K.~Slevin, and T.~Ohtsuki: Phys. Rev. Lett. {\bfseries 89} (2002)
  256601.

\bibitem{Asada2004}
Y.~Asada, K.~Slevin, and T.~Ohtsuki: Phys. Rev. B {\bfseries 70} (2004) 035115.

\end{thebibliography}

\end{document}